\documentclass[prl,twocolumn,showpacs,amsmath,letter]{revtex4}
\usepackage{graphicx, verbatim}

\newcommand{\Journal}[4]{#1 \textbf{#2}, #3 (#4)}

\begin{document}

\title{Temperature-dependent proximity magnetism in Pt}

\author{W.L. Lim}
\author{N. Ebrahim-Zadeh}
\author{H.G.E. Hentschel} 
\author{S. Urazhdin}
\affiliation{Department of Physics, Emory University, Atlanta, GA 30322}


\begin{abstract}

We experimentally demonstrate the existence of magnetic coupling between two ferromagnets separated by a thin Pt layer. The coupling remains ferromagnetic regardless of the Pt thickness, and exhibits a significant dependence on temperature. Therefore, it cannot be explained by the established mechanisms of magnetic coupling across nonmagnetic spacers. We show that the experimental results are consistent with the presence of magnetism induced in Pt in proximity to ferromagnets, in direct analogy to the well-known proximity effects in superconductivity. 
\end{abstract}
\maketitle

For centuries, platinum metal has been highly valued for its luster and rarity. With the technological revolution, Pt has found many applications owing to its high chemical inertness and catalytic properties. Recently, it was shown  that one can modify the dynamical magnetic characteristics of ferromagnets (F) by passing an electrical current through Pt/F bilayers~\cite{Ando_SHE,Liu_SHE,Demidov_SHE}.  This phenomenon caused by the spin Hall effect in Pt~\cite{spinHall} has generated a significant interest in spin-dependent electronic properties of this material.

Spin Hall effect in Pt originates from the spin-orbit interaction of the predominantly d-type conduction electrons. Because of the large density of states of the d-electrons, Pt also almost satisfies the Stoner criterion for the onset of ferromagnetic ordering~\cite{Stoner}. Indeed, magnetism has been predicted and experimentally observed in Pt nanoparticles~\cite{Xiao_clusters,Liu_clusters,Sakamoto_nanoparticles}, nanocontacts~\cite{Smogunov_nanocontacts}, nanowires~\cite{Delin_nanowires, Teng_nanowires}, and thin films~\cite{Niklasson_films,Thapa_films}. These observations raise a question important for spintronic applications: What are the magnetic properties of thin Pt films in heterostructures with magnetic materials?

We address this question by studying the magnetic properties of structures that consist of two ferromagnets separated by a Pt layer. We observe ferromagnetic coupling between the magnetic layers that depends both on the Pt thickness $d$ and on temperature $T$. The coupling decreases with increasing $T$. It persists up to room temperature RT=$295$~K for sufficently small $d$, but vanishes below $RT$ for larger $d$. These characteristics distinguish this coupling from the well-known  Ruderman-Kittel-Kasuya-Yosida (RKKY) mechanism of coupling between ferromagnets separated by a nonmagnetic spacer, which oscillates with spacer thickness and is generally independent of temperature~\cite{RKKY}. Our analysis indicates that the observed coupling  originates from the magnetization induced in Pt at the interfaces with the ferromagnets, suggesting that spin-dependent properties of both Pt and F are mutually affected in Pt/F spintronic heterostructures.

The structure of our samples was Py(8)Pt($d$)Py(3)Ir$_{20}$Mn$_{80}$(8)Ta(1.5), where thicknesses are in nanometers, and Py=Ni$_{80}$Fe$_{20}$. The thickness $d$ of Pt was varied between $2.0$~nm and $4.0$~nm in increments of $0.4$~nm. The multilayers were deposited at RT on oxidized Si substrates by magnetron sputtering at ultrapure Ar gas pressure of $5$~mTorr. The residual gas pressure was $7\times 10^{-9}$~Torr, as verified by a residual gas analyzer. Substrate roughness was less than $0.5$~nm, as verified by atomic force microscopy. The thicknesses of the deposited layers were monitored by a quartz crystal sensor, which was independently calibrated to a relative precision of better than $5\%$ by low-angle x-ray diffraction measurements of single-layer films.

Our heterostructure formed a giant magnetoresistance (GMR) pseudo-spin valve~\cite{GMR}, where Pt was the nonmagnetic spacer, Py(8) was the free magnetic layer, and Py(3) was magnetically pinned by the exchange-bias~\cite{eb} produced by the antiferromagnet Ir$_{20}$Mn$_{80}$. The top Ta(1.5) served as a protective capping layer. The effective exchange bias field $H_B$ was approximately $400$~Oe at RT. It increased approximately linearly with decreasing temperature, reaching about $2.0$~kOe at $5$~K. The direction of $H_B$ was established during the multilayer deposition by applying a uniform field $H=100$~Oe parallel to the sample surface. $H>0$ was in the direction of $H_B$.  In all the measurements presented below, $H$ was significantly smaller than $H_B$, and therefore its effects on the pinned layer were small.

\begin{figure}
\includegraphics[width=2.5in]{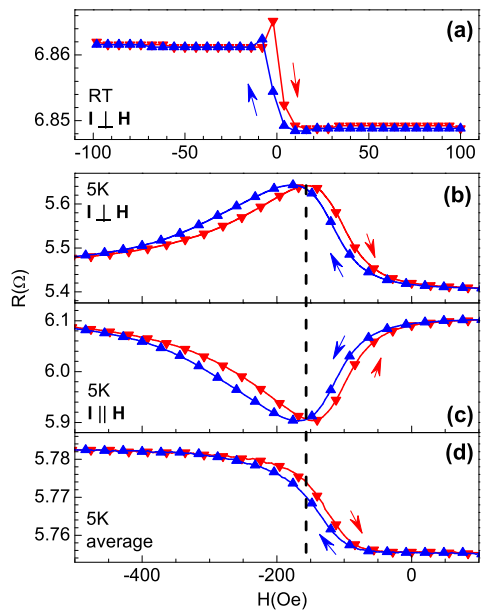}
\caption{\label{fig:fig1} (Color online) $R$ {\it vs} $H$ for the sample with Pt thickness $d=2.8$~nm: (a) at $295$~K, $I\perp H$, (b) at $5$~K, $I\perp H$, (c) at $5$~K, $I\parallel H$, (d) average of (b) and (c). Arrows show the direction of the field scan. Dashed vertical line marks the coupling field $H_E=156$~Oe at $5K$ determined from the data.}
\end{figure}

The magnetic configuration was determined by lock-in measurements of the sheet resistance $R$ in the four-probe Van der Pauw geometry. Figure~\ref{fig:fig1}(a) shows a room-temperature $R$ {\it vs} $H$ curve for the sample with Pt thickness $d=2.8$~nm. At $H>0$, the magnetizations of both layers were oriented parallel to $H$ and $H_B$, resulting in an approximately constant resistance. At RT, the resistance exhibited an abrupt step at small $H$ associated with the reversal of the free Py(8) layer.  The value of $R$ in the antiparallel (AP) configuration of the magnetic layers at $H<0$ was larger than in the parallel (P) configuration at $H>0$, as expected for the GMR effect~\cite{GMR}. When a large $H<0$ was applied, we observed a reduction of $R$ consistent with the reversal of the pinned Py(3) layer into the P configuration with the free layer (not shown).

The shape of the magnetoresistance curves changed as the temperature was reduced. The $R$ {\it vs} $H$ curve acquired with current flowing perpendicular to the field developed a large peak that shifted to increasingly negative $H$ with decreasing temperature, as illustrated for $T=5$~K in Fig.~\ref{fig:fig1}(b). Meanwhile, a similar curve acquired with the current collinear with $H$ exhibited a dip at the same value of $H$ [Fig.~\ref{fig:fig1}(c)]. To confirm that the observed temperature-dependent features are specific to structures with a Pt spacer, we separately fabricated similar heterostructures in which Pt was replaced with a Cu spacer of the same thickness. These samples exhibited behaviors independent of $T$, similar to those shown in Fig.~\ref{fig:fig1}(a).

Since the presence of peaks/dips in the magnetoresistance is controlled by the direction of current relative to the magnetization, these features can be attributed to the anisotropic magnetoresistance (AMR) of the free Py(8) layer. Indeed, the sample resistance exhibited a sinusoidal $180^\circ$-periodic dependence on the direction of $H$ consistent with AMR~\cite{O'Handley},  superimposed on the $360^\circ$-periodic variation due to GMR. By averaging the data for the two orthogonal directions of current, we eliminated the contribution of AMR and obtained GMR steps similar to those observed at RT, as illustrated in Fig.~\ref{fig:fig1}(d). The resistance reached a midpoint between its values in the P and AP state at the same $H$ as the peaks/dips, as illustrated by a dashed vertical line.

Based on these observations, we conclude that the reversal mechanism of the Py(8) layer changes as the temperature is lowered. The absence of AMR peaks associated with the reversal of the free Py(8) layer at RT indicates that the reversal occurs abruptly, likely via a domain wall sweeping across this layer. In contrast, the large amplitude of the AMR peaks at lower temperatures indicates that the reversal occurs by gradual rotation of magnetization.  

Since at low temperatures the reversal peaks are also shifted to incrasingly larger $H<0$, such behaviors can be attributed to a competition between $H$ and an effective field $H_E$ caused by ferromagnetic coupling to the pinned Py(3) layer. The two fields are not exactly collinear with each other, resulting in a rotational reversal of the magnetization and consequently a large associated AMR signal.

\begin{figure}
\includegraphics[width=2.6in]{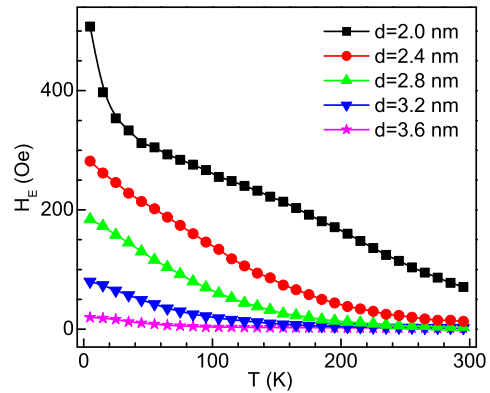}
\caption{\label{fig:fig2} (Color online) Measured temperature dependence of the effective exchange coupling field $H_E$ acting on the Py(8) layer for samples with different thicknesses $d$ of the Pt spacer, as labeled.}
\end{figure}

We determined the magnitude of $H_E$ by averaging the values of $H$ corresponding to AMR peaks obtained with increasing and decreasing field [Figs.~\ref{fig:fig1}(b,c)]. We also verified that, when the AMR peaks were absent, the GMR steps were symmetric with respect to $H$, and consequently $H_E$ was negligible. Figure~\ref{fig:fig2} summarizes the obtained dependencies of $H_E$ on temperature for several samples with different thicknesses $d$ of the Pt spacer. All samples with $d<4$~nm exhibited ferromagnetic coupling between the Py layers at the lowest experimental temperature of $5$~K. At large $d$, the coupling was small and rapidly disappeared with increasing $T$. The coupling increased with decreasing $d$, and persisted to RT in samples with $d\le 2.4$~nm.

Magnetic coupling has been observed in magnetic heterostructures incorporating a variety of nonmagnetic spacers such as Cr, Cu, and Ru~\cite{Grunberg,Parkin}. 
This coupling exhibited typical characteristics of the RKKY interaction mechanism~\cite{RKKY,Slonczewski}: it was oscillatory with respect to the spacer thickness and was generally independent of temperature. Besides RKKY coupling, ferromagnetic layers separated by a spacer can also become coupled due to roughness and/or alloying at the interfaces, resulting in their direct exchange interaction through "pinholes"~\cite{Dieny}. This coupling is also generally independent of temperature.

The observed coupling through thin Pt spacers is strongly temperature dependent and remains ferromagnetic regardless of the spacer thickness, as illustrated in Fig.~\ref{fig:fig2}. Based on these observations, we conclude that this coupling is not caused by the RKKY mechanism or the pinholes but is instead associated with the thermodynamic magnetic properties of Pt. This hypothesis is supported by both the experimental and the theoretical evidence for the existence of magnetism in confined Pt structures~\cite{Xiao_clusters,Smogunov_nanocontacts,Liu_clusters,
Sakamoto_nanoparticles,Delin_nanowires,Teng_nanowires,Niklasson_films,Thapa_films}.

The thermodynamic magnetic properties of the Pt layer can be described by the phenomenological free energy density function~\cite{Landau,Doniach}
\begin{equation}\label{eq:f}
f(x,T)=\frac{\alpha(T)}{2}{\mathbf M}^2+\frac{\gamma(T)}{2}\left[\frac{d{\mathbf M}}{dx}\right]^2-{\mathbf M}\cdot {\mathbf H},
\end{equation}
where the first term describes the spin polarizability of Pt, the second term describes the spatial spin correlations, and the last term is the Zeeman energy.
In this expression, $x$ is the direction normal to the layers, ${\mathbf M}$ is the local magnetization in Pt, $\alpha(T)$ is the inverse susceptibility of Pt with respect to the effective exchange field, and $\gamma(T)$ is the 
exchange stiffness of Pt. A precise quantitative treatment of magnetic susceptibility may require higher orders of ${\mathbf M}$ in the expression for the free energy. For simplicity we neglect such nonlinear effects. We also neglect the Zeeman term in Eq.~(\ref{eq:f}), since $H$ was small compared to the typical exchange fields. 

Exchange interaction at the interfaces with Py can be described by a local effective field $H_{ex}=JM_0\approx 10^4$~Tesla, where $J$ is the exchange coupling across the interface, and $M_0$ is the magnetization of Py. At the interface, the exchange field dominates over other contributions to the free energy density Eq.~(\ref{eq:f}). Assuming that the Pt layer is located between $x=-d/2$ and $d/2$,
$$
M(x=\pm d/2)=JM_0/2\alpha.
$$
We consider only the P/AP configurations, where the magnetizations of both ferromagnets are collinear with the z-axis. It is clear from Eq.~(\ref{eq:f}) that the free energy is minimized when $M_x(x)=M_y(x)=0$, $M\equiv M_z$. The energy per unit surface area of the Pt layer is
\begin{equation}\label{eq:F}
F =\int_{-d/2}^{d/2}[\frac{\alpha}{2} M^2 + \frac{\gamma}{2} \left(\frac{dM}{dx}\right)^2]dx.
\end{equation}

Minimizing this functional with respect to $M(x)$, we obtain
$$
M(x)=\frac{JM_0}{\alpha}\frac{\exp[(x/\xi)] \pm \exp[(-x/\xi)]}
{\exp[(d/2\xi)] \pm \exp[(-d/2\xi)]},
$$
where the $+(-)$ sign corresponds to the P(AP) configuration of the magnetic layers, and $\xi= \sqrt{\gamma/\alpha}$ is the characteristic decay length of the magnetization in Pt away from the interface. This parameter is analogous to the coherence length introduced in the theory of proximity superconductivity~\cite{sc_proximity}. Inserting $M(x)$ back into the expression Eq.~(\ref{eq:F}) for the energy $F_{P,AP}$ in P and AP configurations, we obtain
\begin{equation}\label{eq:FPAP}
F_{P,AP}=\frac{J^2M_0^2\xi^3}{\gamma}\frac{\exp[(d/2\xi)] \mp \exp[(-d/2\xi)]}{\exp[(d/2\xi)] \pm \exp[(-d/2\xi)]}.
\end{equation}

To relate our analysis to the data, we note that the change of the Zeeman energy of the free Py layer when it reverses between AP and P configurations is 
\begin{equation}\label{eq:dF}
2M_0t_{free}H_E=F_{AP}-F_P=\Delta F,
\end{equation}
where $t_{free}=8$~nm is the thickness of the free layer. From Eqs.~(\ref{eq:FPAP}), (\ref{eq:dF}),
\begin{equation}\label{eq:he}
H_E= \frac{J^2M_0\xi^3}{\gamma t_{free}\sinh[(d/\xi)]}.
\end{equation}

\begin{figure}
\includegraphics[width=\columnwidth]{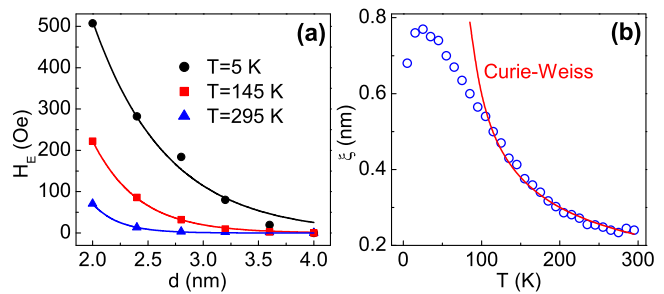}
\caption{\label{fig:fig3} (Color online) (a) Symbols: $H_E$ {\it vs} $d$ at $T=5$~K (circles), $145$~K (squares), and $295$~K (triangles). Curves: results of fitting with Eq.~\ref{eq:he}. (b) Symbols: $\xi$ {\it vs} $T$ determined  from the fits as shown in panel (a). Curve: fitting of the $T>100$~K data with the Curie-Weiss dependence, as described in the text.}
\end{figure}

At $T>100$~K, Eq.~(\ref{eq:he}) provided an excellent description for the experimental dependence of $H_E$ on the Pt thickness $d$, as illustrated in Fig.~\ref{fig:fig3}(a) for $T=145$~K and $295$~K. At $T<100$~K, the values of $H_E$ at large $d$ fell below the dependence predicted by Eq.~(\ref{eq:he}). 
These behaviors are consistent with larger induced magnetization in Pt, resulting in increasingly nonlinear responce to the effective exchange field and consquently more significant deviations from our linear-responce analysis.

Figure~\ref{fig:fig3}(b) shows the dependence of $\xi$ on temperature obtained by fitting the $H_E$ {\it vs} $d$ data as shown in Figure~\ref{fig:fig3}(a).
At high temperature, the coherence length increases nonlinearly with decreasing $T$. It falls onto an approximately linear dependence between $50$~K and $150$~K, and  decreases below $25$~K. We can compare these behaviors to the dependence expected from the Curie-Weiss law~\cite{kittel}, whereby $\alpha$ varies approximately linearly with $T$~\cite{Landau}. Neglecting the dependence of $\gamma$ on temperature, we obtain $\xi(T)\approx\xi_0/\sqrt{(T/T_0-1)}$. Here, $\xi_0$ is independent of $T$, and $T_0$ has units of temperature and can be interpreted as the effective Curie temperature of the Pt film over the applicable temperature range. This formula  provides a good fitting to the data at $T>100$~K, as shown by a curve in Figure~\ref{fig:fig3}(b). However, at lower $T$ the data fall significantly below the Curie-Weiss dependence. The deviations of $H_E(d,T)$ from the dependence predicted by Eq.~(\ref{eq:he}) also become significant at $T<100$~K. These observations point to the emergence of new nonlinear physical mechanisms that limit the magnetic coherence length at low temperatures. Future theoretical analyses will likely elucidate the origin of these behaviors.

It is possible that the properties of Pt in our structures are affected by magnetic impurities inevitably present in the source material. We can estimate  the effects of such impurities in comparison to the proximity effects discussed above. The effects of interfaces with ferromagnets on a $3$~nm-thick Pt layer are at least as significant as those of a series of atomic magnetic impurity layers with a $3$~nm spacing. The equivalent volume density of such impuruties is $15$ at.~$\%$, much larger than the typical concentrations of impurities in the deposition sources. 

Confinement effects can also induce surface magnetism in Pt films even without adjacent ferromagnets~\cite{Niklasson_films,Thapa_films}, contributing to the observed proximity magnetism. It will be possible to determine the significance of these contributions, for example, by analyzing the dependence of magnetic coupling through Pt on the magnetization of the adjacent ferromagnets, and comparing to the dependence on $M_0$ predicted by Eq.~(\ref{eq:he}).

Our observation of proximity-induced magnetism in Pt has a number of significant implications for spintronics. First, the contribution of Pt can significantly affect the static and the dynamic magnetic properties of nanostructures. Second, spin-dependent transport in Pt/F heterostructures may be affected not only by the spin Hall effect but also by spin-dependent scattering within the magnetized Pt volume. In particular, the effective spin-diffusion length in thin Pt layers should depend on the magnetic environment. In spin-transfer~\cite{Slonczewski96} and spin-pumping~\cite{spinpumping} structures, electron spins noncollinear with the magnetization are important. Larmor spin precession in the proximity-magnetized Pt can dramatically affect these spins. In all cases, the expected distinguishing feature of the effects of induced magnetism in Pt is a strong dependence of these effects on temperature. 
 
This work was supported in part by the NSF Grant DMR-0747609.

\end{document}